\documentclass[preprint,aps]{revtex4}
\usepackage{epsfig}
\begin{document}
\title{STABILITY EQUATION AND TWO-COMPONENT EIGENMODE FOR DOMAIN WALLS
IN A SCALAR POTENTIAL MODEL}
\author{G. S. Dias, E. L. Gra\c{c}a
and R. de Lima Rodrigues \\
Centro Brasileiro de Pesquisas F\'\i sicas\\
Rua Dr. Xavier Sigaud, 150\\
CEP 22290-180, Rio de Janeiro-RJ, Brazil }

\begin{abstract}
The connection between the supersymmetric quantum mechanics
involving two-component eigenfunctions and the stability equation
associated with two classical configurations is investigated and a
matrix superpotential is deduced. The question of stability is
ensured for the Bogomol'nyi-Prasad-Sommerfield (BPS) states on two
domain walls in a scalar potential model containing up to
fourth-order powers in the fields, which is explicit demonstrated
using the intertwining operators in terms of two-by-two matrix
superpotential in the algebraic framework of supersymmetry in
quantum mechanics. Also, a non-BPS state is found to be non-stable
via the fluctuation Hessian matrix.

\vspace{1.0cm}


\vspace{1cm} {\small Permanent address: To RLR is Unidade
Acad\^emica de Ci\^encias Exatas e da Natureza, Universidade
Federal de Campina Grande, Cajazeiras -- PB, 58.900-000, Brazil.
To GSD is Centro Federal de Educa\c{c}\~ao Tecnol\'ogica do
Espírito Santo, Unidade de Vit\'oria, Jucutuquara, CEP 29040-780,
Vit\'oria-ES. To ELG is Departamento de F\'\i sica, Universidade
Federal Rural do Rio de Janeiro. Antiga Rodovia Rio-S\~ao Paulo Km
47, BR 465. CEP 23.890-000, Serop\'edica-RJ. E-mails to RLR
rafaelr@cbpf.br or rafael@df.ufcg.edu.br. E-mail for GSD is
gilmar@cefetes.br.}

\vspace{0.5cm} {\small To appear in International Journal of Modern
Physics A, Vol 21 (2006), hep-th/0205195.}
\end{abstract}

\maketitle



\section{Introduction}

\paragraph*{}

The algebraic framework of Supersymmetry in Quantum Mechanics
(SUSY QM), as formulated by Witten \cite{W,P}, may be elaborated
from a 2-dimensional model. The SUSY QM generalization of the
harmonic oscillator raising and lowering operators has been
several applications \cite{Fred,G,La,C,D}. The generalization of
SUSY QM for the case of matrix superpotential, is well known in
the literature for a long time. See, for example, for
one-dimensional systems the works in Ref.
\cite{Ca90,CI,CF2,ACIN,Fu,R01}.

 Recently, those
has been investigated the superpotential associated with the
linear classical stability from the static solutions for  systems
of two real scalar fields in (1+1) dimensions that present powers
up to sixth-order \cite{RPV98} and one field \cite{vvr02}. In the
case of two coupled scalar fields,  the static field
configurations were determined via Rajaraman's trial orbit method
\cite{Raja,Raja2}. However, while Rajaraman has applied your
method for the equation of motion, here one uses the trial orbit
method for the first order differential equations
\cite{RPV98,BWLR}.

Also, recently, the reconstruction of a single real scalar field
theory \cite{vvr02} from excitation spectra of inflation potential
in framework of inflationary cosmology with the production of a
topological defect has been implemented the via superpotentials
\cite{vachas04}, and a superfield formulation of the central
charge anomaly in quantum corrections to soliton solutions with
N=1 SUSY has been investigated \cite{KS04}.

In this work, for the time being, we will only our attention
focous on SUSY QM for the two-component eigenmodes of the
fluctuation operator; of course, for a zero mode we have a
corresponding $\Psi_0(z)=\left(
\begin{array}{c}\eta_0 \\ \xi_0 \end{array}\right).$ Nevertheless,
this is not necessarily true for an arbitrary non-trivial real
eigenvalue $\omega^2_n$. Here, we shall determine the number of
bound states. We also construct the two-by-two matrix
superpotential for SUSY QM in the case of the stability equation
associated with 2-dimensional potential model considered in
literaure \cite{Shif99}. We consider the classical configurations
with domain wall solutions, which are bidimensional structures in
3+1 dimensions. They are static, non-singular, classically stable
Bogomol'nyi \cite{Bogo} and Prasad-Sommerfield \cite{PS} (BPS)
soliton (defect). Also, the non-BPS defect solution to field
equations with finite localized energy associated with a real
scalar field potential model is found.

The BPS states are classical configurations that satisfy the first
order differential equations and the second order differential
equations (equations of motion). On the other hand, non-BPS defect
satisfies the equation of motion, but does not obey the first
order differential equations.

Relativistic systems with topological defect appear to be extended
objects such as strings or membranes, for which have been obtained
a necessary condition for some stable stringlike intersections,
providing the marginal stability curves \cite{at91}. Recently,
marginal stability and the metamorphosis of BPS states have been
investigated \cite{Shif01b}, the via SUSY QM, with a detailed
analysis for a 2-dimensional $N=2-$Wess-Zumino model in terms of
two chiral superfields, and composite dyons in
$N=2$-supersymmetric gauge theories.

Domain walls have been recently exploited in a context that
stresses their connection with  BPS-bound states \cite{guila02}.
Let us point out that some investigations are interesting in
connection with Condensed Matter, Cosmology, coupled field
theories with soliton solutions
\cite{Morris,Morris2,carro,safin99,dani00,ShifV} and one-loop
quantum corrections to soliton energies and central charges in the
supersymmetric $\phi^4$ and sine-Gordon models in (1+1)-dimensions
\cite{Graha,rebhan99}. The work of Ref. \cite{Graha} reproduces
the results for the quantum
 mass of the SUSY solitons previously obtained in Ref. \cite{rebhan99}.
 Recently, the reconstruction of
2-dimensional scalar field potential models has been considered,
and quantum corrections to the solitonic sectors of both
potentials are pointed out \cite{GN}. The quantization of
two-dimensional supersymmetric solitons is in fact
 a surprisingly intricate issue in many aspects
\cite{rebhan97,goldha01,nieuwen01,wimmer02,izquierdo02,Iz2}.

In the paper work, a connection between SUSY QM  developments and
the description of such a physical system with stability equation
is expressed in terms of two-component wave functions. This leads
to four-by-four supercharges and supersymmetric Hamiltonian
matrices whose bosonic sectors possesses a fluctuation operator
$(O_F)$ associated with two-component eigenstates in terms of BPS
states.

This paper is organized as follows: In Section II, we investigate
domain walls configurations for two coupled scalar fields;
supersymmetric non-relativistic quantum mechanics with
two-component wave functions is implemented in Section III. In
Section IV, from a scalar potential model of two coupled scalar
fields we analyze the stability of non-BPS and BPS states, and a
new matrix superpotential to supersymmetric non-relativistic
quantum mechanics with two-component wave functions is also
implemented. Our Conclusions are presented in Section V.

\section{Domain walls from two coupled scalar fields}

\paragraph*{}

In this section, we investigate a potential model in terms of two
coupled real scalar fields in (1+1) dimensions that present
classical soliton solutions known as domain walls.

The Lagrangian density for such a non-linear system, in natural
units, is written as

\begin{equation}
\label{E18} {\cal L}\left(\phi, \chi, \partial_{\mu} \phi,
\partial_{\mu}\chi\right)
= \frac{1}{2}\left(\partial_{\mu}\phi\right)^2+
\frac{1}{2}\left(\partial_{\mu}\chi\right)^2 -V(\phi, \chi),
\end{equation}
where $\eta^{\mu\nu}=diag(+, -)$ is the metric tensor. Here, the
potential $V=V(\phi,\chi)$ is any positive semidefinite function of
$\phi$ and $\chi$, which can be written as a sum of perfect squares
and must have at least two different zeros in order have present
domain walls as possible solutions. The general classical
configurations obey the equations:

\begin{equation}
\label{E19}
 {\sqcup\!\! \!\!\sqcap}\phi+\frac{\partial }{\partial \phi}V=0, \qquad
{\sqcup\!\! \!\!\sqcap}\chi +\frac{\partial }{\partial \chi}V=0.
\end{equation}
For static soliton solutions, the equations of motion become the
following system of non-linear differential equations:

\begin{eqnarray}
\label{eso}
 \phi^{\prime\prime}&&=\frac{\partial }{\partial \phi}V
\nonumber \\
 \chi^{\prime\prime}&&=\frac{\partial }{\partial \chi}V,
\end{eqnarray}
where primes stand for differentiations with respect to the space
variable. There appears in the literature a trial orbit method for
the attainment of static solutions for certain positive
potentials. This method yields, at best, some solutions to Eq.
({\ref{eso}) and by no means to all classes of potentials
\cite{Raja}. In this work, the trial orbit method has been applied
to systems of two coupled scalar fields containing up to
fourth-order powers in the fields.

If the potential is a sum of perfect squares given by

\begin{equation}
\label{PQ}
2V(\phi,\chi)=\left(\frac{\partial}{\partial \phi}
W\right)^2+\left(\frac{\partial}{\partial \chi}W\right)^2
\end{equation}
one can deform the total energy

\begin{equation}
\label{ETG} E=\int_{-\infty}^{+\infty} dz\frac
12\left[\left(\phi^{\prime}\right)^2 +\left(\chi^{\prime}\right)^2
+ 2V(\phi,\chi)\right],
\end{equation}
under the  BPS form of the energy, consisting of a sum of squares
and surface terms,

\begin{equation}
\label{ET} E=\int_{-\infty}^{+\infty} dz\left[\frac
12\left(\phi^{\prime}- \frac{\partial}{\partial \phi}
W\right)^2+\frac 12\left(\chi^{\prime}- \frac{\partial}{\partial
\chi}W\right)^2 +\frac{\partial}{\partial z}W \right]
\end{equation}
so that the first and second terms are always positive. In this
case the lower bound of the energy (or classical mass) is given by
the third term, viz.,

\begin{equation}
\label{Ebogo}
E\geq\left|\int_{-\infty}^{+\infty} dz
\frac{\partial}{\partial z}W[\phi(z), \chi(z)]\right|,
\end{equation}
where the superpotential $W=W[\phi(z), \chi(z)]$ shall be
discussed below.

The BPS mass bound of the energy result in a topological charge
and is given by $W_{ij}=W[M_j]-W[M_i],$ where $M_i$ and $M_j$
represent the vacuum states. It is required that $\phi$ and $\chi$
satisfy the BPS state conditions \cite{Bogo}:

\begin{eqnarray}
\label{E23}
  \phi^{\prime}&&=  \frac{\partial W}{\partial\phi},  \nonumber\\
\chi^{\prime}&&=\frac{\partial W}{\partial\chi}.
\end{eqnarray}
Note that the BPS states saturate the lower bound so that
$E_{BPS}=T_{ij}=|W_{ij}|,$ where $W_{ij}$ is the central charge of
the realization of $N=1$ SUSY in 1+1 dimensions.

\section{SUSY QM AND LINEAR STABILITY}

\paragraph*{}

In this section, we present the connection between the SUSY QM and
the stability of classical static domain walls against small
quantum fluctuations, in which for BPS domain walls, the
fluctuation Hessian is the bosonic part of a SUSY matrix
fluctuation operator, and thus, it is positive definite.

 Now, let us analyze the classical stability of the domain walls
in the non-linear system considered in this work, by considering
small perturbations around $\phi (z)$ and $\chi (z)$:

\begin{equation}
\label{E32}
  \phi(z,t)=\phi(z)+\eta(z,t)
\end{equation}
and

\begin{equation}
\label{E33}
  \chi(z,t)=\chi(z)+\xi(z,t).
\end{equation}
Next, let us expand the fluctuations $\eta(z,t)$ and $\xi(z,t)$ in
terms of normal modes:

\begin{equation}
\label{E34}
\eta (z,t) = \sum_n \epsilon_n \eta_n (z) e^{i\omega_n
t}
\end{equation}
and

\begin{equation}
\label{E35} \xi (z,t) = \sum_n c_n \xi_n (z) e^{i\omega_n t},
\end{equation}
where $\epsilon_n$ and $c_n$ are chosen so that $\eta$ and $\chi$
are real (there are scattering states). Thus, considering a Taylor
expansion of the potential $V(\phi, \chi)$ in powers of $\eta$ and
$\xi$, by retaining the first order terms in the equations of
motion (\ref{E19}), one gets a Schr\"odinger-like equation for
two-component wave functions

\begin{equation}
\label{EE} O_F \Psi_{n} = \omega_n^2\Psi_{n}, \quad
\Psi_{n}=\left(
\begin{array}{cc}
\eta_n(z) \\
 \xi_n(z)
\end{array}\right), \quad n=0, 1, 2, \cdots_.
\end{equation}

The frequency terms, $\omega_n^2,$ arise from the 2-nd times
derivatives and the matrix fluctuation operator becomes

\begin{equation}
\label{E37} O_F=\left(
\begin{array}{cc}
-\frac{d^2}{dz^2} +\frac{\partial^2}{\partial\phi^2}V &
\frac{\partial^2}{\partial \chi\partial\phi}V \\
 \frac{\partial^2}{\partial \phi\partial\chi}V &
-\frac{d^2}{dz^2} +\frac{\partial^2}{\partial\chi^2}V
\end{array}\right)_{\vert\phi =\phi(z),
\chi =\chi(z) }\equiv-{\bf I}\frac{d^2}{dz^2}+V_F(z),
\end{equation}
where {\bf I} is the two-by-two identity matrix and

\begin{eqnarray}
\label{DP} V_{F12}(z)=V_{F21}(z)\equiv\frac{\partial^2}{\partial
\chi\partial\phi}V&&=\frac{\partial^2}{\partial\phi\partial\chi}V,\nonumber\\
V_{F11}(z)&&\equiv\frac{\partial^2}{\partial\phi^2}V\nonumber\\
V_{F22}(z)&&\equiv\frac{\partial^2}{\partial\chi^2}V.
\end{eqnarray}
are the elements of fluctuation Hessian matrix, $V_F(z).$ Thus,
the eigenvalues of the Hessian matrix at each stationary point
$M_i$ provide a classification of gradient curves of the
superpotential.

We can get the masses of the bosonic particles, using the results
above, from the second derivatives of the potential:

\begin{eqnarray}
\label{massa}
  m^2_\phi\equiv
\frac{\partial^2V}{\partial\phi^2}\left|_{z\rightarrow\pm\infty}\right.
\nonumber\\
  m^2_\chi\equiv
\frac{\partial^2V}{\partial\chi^2}\left|_{z\rightarrow\pm\infty}\right..
\end{eqnarray}

Coming to the case of a single real scalar field \cite{vvr02}, we
can realize, {\it a priori}, the two-by-two matrix superpotential
that satisfies the following Ricatti equation associated to the
non-diagonal flutuaction Hessian $V_{F}(z)$ given by Eq.
(\ref{E37}):

\begin{equation}
\label{ER} \hbox{{\bf W}}^2(z)+\hbox{{\bf
W}}^{\prime}(z)=V_{F}(z)=\left(
\begin{array}{cc}
 V_{F11}(z) &  V_{F12}(z)\\
 V_{F12}(z) & V_{F22}(z)
\end{array}\right)_{\vert\phi=\phi(z),\chi=\chi(z)},
\end{equation}
whose solution is a Hermitian operator

\begin{equation}
\label{spqm} \hbox{{\bf W}}(z)=\left(
\begin{array}{cc}
 f(\phi,\chi) & g(\phi,\chi) \\
 g(\phi,\chi) & h(\phi,\chi)
\end{array}\right)_{\vert\phi=\phi(z),\chi=\chi(z)}={\hbox{{\bf W}}}^{\dagger},
\end{equation}
where we must use the BPS state conditions (\ref{E23}) and
$V_{Fij}(z)$ is given by the fluctuation Hessian. The Eqs.
(\ref{ER}) and (\ref{spqm}) are correct only for BPS states. Thus
the bosonic zero-mode

\begin{equation}
\label{zmp}
\Psi_-^{(0)}(z)=\Psi_{0}(z)=N\left(
\begin{array}{cc}
\eta_0(z) \\
 \xi_0(z)
\end{array}\right), \quad
\eta_0=\frac{d\phi}{dz}, \quad \xi_0=\frac{d\chi}{dz},
\end{equation}
satisfies the annihilation condition $A^-\Psi_-^{(0)}(z)=0,$ i.e.,
$\frac{d}{dz}\Psi_-^{(0)}(z)= \hbox{{\bf W}}\Psi_-^{(0)}(z),$
where $N$ is the normalization constant. According to the Witten'
SUSY model \cite{W,Fred,R01}, we have

\begin{equation}
\label{A+-}
{\cal A}^{\pm}=\pm\hbox{{\bf I}}
\frac{d}{dz}+\hbox{{\bf W}}(z), \quad
\Psi_{\hbox{SUSY}}^{(n)}(z)=\left(
\begin{array}{cc}
\Psi_-^{(n)}(z) \\
\Psi_+^{(n)}(z)
\end{array}\right),
\end{equation}
where  $\Psi_{\pm}^{(n)}(z)$ are two-component eigenfunctions. In
this case, the graded Lie algebra of the supersymmetry in quantum
mechanics for the BPS states may be readily realized as

\begin{equation}
\label{E40} H_{SUSY} = [Q_- ,Q_+ ]_+ = \left(
\begin{array}{cc}
{\cal A}^+ {\cal A}^- & 0 \\
0 & {\cal A}^- {\cal A}^+
\end{array}\right)_{4\hbox{x}4}= \left(
\begin{array}{cc}
{\cal H}_- & 0 \\
0 & {\cal H}_+
\end{array}\right),
\end{equation}

\begin{equation}
\label{41}
 \left[H_{SUSY} , Q_{\pm}\right]_- = 0 =
(Q_-)^2=(Q_+)^2,
\end{equation}
where $Q_{\pm}$ are the 4 by 4 supercharges of Witten' $N=2$ SUSY
model, viz.,

\begin{equation}
\label{SC} Q_- = \sigma_-\otimes{\cal A}^-, \quad
Q_+=Q_-^{\dagger}= \left(
\begin{array}{cc}
0 & {\cal A}^+ \\
0 & 0
\end{array}\right)=\sigma_+\otimes{\cal A}^+,
\end{equation}
with the intertwining operators, ${\cal A}^{\pm}$ are given in
terms of two-by-two matrix superpotential by Eq. (\ref{A+-}) and
$\sigma_{\pm}=\frac 12(\sigma_1\pm i\sigma_2),$ where $\sigma_1$
and $\sigma_2$ are Pauli matrices. Of course, the bosonic sector
of $H_{SUSY}$ is exactly the fluctuation operator given by ${\cal
H}_-=O_F=-\frac{{\bf I}d^2}{dz^2} +{\bf V}_{F}(z),$ where ${\bf
V}_-={\bf V}_{F}(z)$ is the non-diagonal fluctuation Hessian. The
SUSY partner of ${\cal H}_-$ is ${\cal H}_+=-\frac{{\bf
I}d^2}{dz^2} +{\bf V}_{+}(z),$ where $V_+=V_F(z)-W^{\prime}.$

Thus the $N$=2 SUSY  algebra in (1+1)-dimensions in terms of real
supercharges becomes

\begin{equation}
\label{sa}
[Q^{i}_{\nu},Q^{j}_{\beta}]=
2\delta^{ij}(\gamma^{\mu}\gamma^0)_{\nu,\beta}P_{\mu}+
2i\left(\gamma^5\gamma^0\right)_{\nu,\beta}W^{ij}, \quad
\nu,\mu,\alpha,\beta=0,1,
\end{equation}
where $P_{\mu}=(P_0,P_1)$ is the energy-momentum operator, and the
Majorama basis for two-by-two $\gamma$-matrices is realized in
terms of the Pauli matrices, $\gamma^0=\sigma_2,
\gamma^1=i\sigma_3$ and $\gamma^5=\gamma^0\gamma^1=-\sigma_1.$ In
this case, the $SO(3,1)$ Lorentz symmetry  in 3+1 dimensions
reduces in 1+1 dimensions to the product of the Lorentz boost in
1+1 and a global symmetry associated with the fermion charge,
viz., $SO(1,1)\otimes U(1)_R.$

In the rest frame $P_1\rightarrow 0$ and $P_0\rightarrow
M=\sqrt{P^{\mu}P_{\mu}},$ the algebra (\ref{sa}) may be rewritten
as

\begin{eqnarray}
\label{sas}
\left(Q^{1}_{1}\right)^2&&=\left(Q^{2}_{2}\right)^2=M+\mid W_{ij}\mid\nonumber\\
\left(Q^{1}_{2}\right)^2&&=\left(Q^{2}_{1}\right)^2=M-\mid
W_{ij}\mid
\end{eqnarray}
with all other anticommutators vanishing, where the supercharge
$W_{ij}$ is the lower bound for the classical mass, $M\geq\mid
W_{ij}\mid,$  so the BPS states saturate the bound states
$E_{BPS}=\mid W_{ij}\mid.$ Also, it is easy to show that the
linear stability is satisfied, i.e.,

\begin{equation}
\label{freq}
 \omega^2_n = \left < O_F\right
>=\left< {\cal A}^+ {\cal A}^-\right >= ({\cal
A}^-\Psi_n)^{\dag}({\cal A}^-\Psi_n) =\mid
{\tilde\Psi}_n\mid^2\geq 0, \quad {\tilde\Psi}_n={\cal A}^-\Psi_n.
\end{equation}
 as it  has been anticipated. Note
that we have set $O_F\equiv{\cal A}^+ {\cal A}^-,$ where the
intertwining operators ${\cal A}^{\pm}$ of SUSY QM must be given
in terms of the matrix superpotential, ${\bf W}(z).$

Therefore, the two-component normal modes in (\ref{EE}) satisfy
$\omega_n{^2} \geq 0$, so that the stability of the BPS domain
wall is ensured, for $\phi\neq 0$ and $\chi\neq 0.$

A realization of SUSY QM model for this system must necessarily be
modified \cite{R01,vachas04}. The essential reason for the
necessity of modification is that the Ricatti equation given by
(\ref{ER}) is reduced to a set of first-order coupled differential
equations. In this case, the superpotential is not necessarily
according to the system described by one-component wave functions
with SUSY in the context of non-relativistic Quantum Mechanics
\cite{W,Fred,G,La,C,D,R01}, which is defined as
$W(x)=\frac{1}{\psi_-^{(0)}}\frac{d}{dx}\psi_-^{(0)}(x),$.

Therefore, as the bosonic zero-mode is associated with a
two-component eigenfunction, $\Psi_-^{(0)}(z)$, one may write the
matrix superpotential only in the form
$\frac{d}{dz}\Psi_-^{(0)}(z)= \hbox{{\bf W}}\Psi_-^{(0)}(z)$
\cite{R01}. Also, we can find the eigenmodes of the supersymmetric
partner ${\cal H}_-$ from those of ${\cal H}_-\equiv O_F$, and the
spectral resolution of the hierarchy of matrix fluctuation
operator may be achieved in an elegant way. In this case, the
intertwining operators ${\cal A}^+({\cal A}^-)$ convert an
eigenfunction of ${\cal H}_-({\cal H}_+)$ to an eigenfunction of
${\cal H}_+({\cal H}_-)$ with the same energy and simultaneously
destroys (creates) a node of
$\Psi_-^{(n+1)}(z)\left(\Psi_+^n(z)\right).$

\section{ The scalar potential model}

Let us now consider a positive potential, $V(\phi, \chi),$ with
the explicit form:

\begin{equation}
V(\phi, \chi)=\frac{1}{2}\lambda^2\left(\phi^2-\frac{m^2}
{\lambda^2}\right)^2+\frac 12\alpha^2\chi^2(\chi^2+4\phi^2)+
\alpha\lambda \chi^2(\phi^2-\frac{m^2} {\lambda^2}).
 \label{EV}
\end{equation}
This potential can be written as a sum of perfect squares. For
static soliton solutions, the equations of motion become the
following system of non-linear differential equations:

\begin{eqnarray}
\label{E20}
 \phi^{\prime\prime}&&=\frac{\partial }{\partial \phi}V
 =2\lambda^2\phi\left(\phi^2-\frac{m^2}{\lambda^2}\right)+
2\alpha\chi^2\phi(2\alpha+\lambda)\nonumber \\
 \chi^{\prime\prime}&&=\frac{\partial }{\partial \chi}V
 =-2\alpha\chi(\frac{m^2}{\lambda}-\lambda\phi^2-
\alpha\chi^2)+4\alpha^2\chi\phi^2.
\end{eqnarray}

The above potential contains only two free parameters, viz.,
$\alpha$ and $\lambda.$ It is non-negative for all real $\alpha$
and $\lambda.$ This potential is of interest because  it has
solutions like BPS and non-BPS, however,  only for $\alpha\lambda
>0,$ does it have 4 minima in which $V=0.$ Note that this
potential has the discrete symmetry: $\phi\rightarrow-\phi$ and
$\chi\rightarrow-\chi$, so that we have a necessary (but
non-sufficient) condition that it must have at least two zeroes in
order that domain walls  can exist.

The superpotential  $W(\Phi, {\mbox{\boldmath $\chi$}})$, as
proposed in the literature, yields the component-field potential
$V(\phi,\chi)$ of Eq. (\ref{EV}), which can be written as

\begin{equation}
 W(\Phi,  {\mbox{\boldmath $\chi$}})=\frac{m^2}{\lambda}\Phi-
\frac{\lambda}{3}\Phi^3- \alpha\Phi {\mbox{\boldmath $\chi$}}^2,
\end{equation}
where $\Phi$ and  ${\mbox{\boldmath $\chi$}}$ are chiral
superfields which, in terms of bosonic ($\phi, \chi$), fermionic
$(\psi,\xi)$ and auxiliary fields $(F, G)$, are $\theta-$expanded
as shown below:

\begin{eqnarray}
\label{SF}
  \Phi &&=\phi+\bar\theta\psi+\frac{\theta\bar\theta}{2}F, \nonumber\\
  {\mbox{\boldmath $\chi$}} &&=\chi+\theta\xi+\frac{\theta\bar\theta}{2}G,
\end{eqnarray}
where $\theta$ and $\bar{\theta}=\theta^*$ are Grassmannian
variables. The superpotential above, with two interacting chiral
superfields, allows for solutions describing string like "domain
Ribbon" defects embedded within the domain wall. It is
energetically favorable for the fermions within the wall to
populate the domain Ribbons \cite{Morris}.

It is required that $\phi$ and $\chi$ satisfy the BPS state
conditions:

\begin{eqnarray}
\label{cbps}
  \phi^{\prime}&&=-\lambda\phi^{2}-\alpha\chi^2+\frac{m^2}{\lambda},
  \nonumber\\
\chi^{\prime}&&=-2\alpha\phi\chi.
\end{eqnarray}

Clearly, only the bosonic part of those superfields are relevant
for our discussion and one should retain only them and the
corresponding part of the $W$ function. Thus, the vacua are
determined by the extrema of the superpotential, so that

\begin{equation}
  \frac{\partial W}{\partial\phi}=0
\end{equation}
and

\begin{equation}
  \frac{\partial W}{\partial\chi}=0
\end{equation}
providing four vacuum states $(\phi, \chi)$ whose values are
listed below in what follows

\begin{eqnarray}
\label{E26}
  M_1&&=\left(-\frac{m}{\lambda},0\right) \nonumber\\
  M_2&&=\left(\frac{m}{\lambda},0\right)\nonumber\\
  M_3&&=\left(0,-\frac{m}{\sqrt{\lambda\alpha}}\right) \nonumber\\
  M_4&&=\left(0,\frac{m}{\sqrt{\lambda\alpha}}\right).
\end{eqnarray}

When the wall $M_{13}$ is stable, the two vacuum states, $M_1$ and
$M_3$, may be adjacent. They are  energy-degenerated with the
energies of the three walls $M_{23}$, $M_{14}$, $M_{24}$. Of
course, from (\ref{E26}), we can see that we have two more
possible domain walls, viz., $M_{12}$  and $M_{34}$, with
different energies. Indeed, in this work, the potential presents a
$Z_2\hbox{x}Z_2$-symmetry, so that one can build some
intersections between the walls \cite{carro,safin99,dani00,ShifV}.

The generalized system, given by Eq. (\ref{cbps}), can be solved
by the trial orbit method developed by Rajaraman \cite{Raja}.
Indeed, using the trial orbit below for the first order
differential equations \cite{RPV98,BWLR}

\begin{equation}
\label{ot}
(\lambda-\beta)\phi^2+\alpha\chi^2=\frac{m^2}{\lambda}-\gamma,
\end{equation}
we obtain a pair of defect solutions given by

\begin{eqnarray}
\label{ps} \phi(z)&&=\sqrt{\frac{\gamma}{\beta}}
\tanh(\sqrt{\beta\gamma}z)\nonumber\\
\chi(z)&&=\pm\frac{1}{\sqrt{\alpha}}\left(\frac{\gamma}{\beta}(\beta-\lambda)
\tanh^2(\sqrt{\beta\gamma}z)+\frac{m^2}{\lambda}-
\gamma\right)^{\frac 12},
\end{eqnarray}
where $\beta$ and $\gamma$ are two real parameters. Note that we
have an elliptical orbit of the BPS domain wall only for the case
$\lambda> \beta$ and $\gamma<\frac{m^2}{\lambda}.$ When
$\lambda-\beta=\alpha$ and $\gamma<\frac{m^2}{\lambda},$ the orbit
becomes a semi-circle.

\subsection{Non-BPS Solutions}

 The non-BPS wall, for $\phi=0,$ is
described by the following equation of motion:

\begin{equation}
\label{NBPS}
\frac{d^2\chi}{dz^2}=-2\alpha\chi\left(\frac{m^2}{\lambda}-
\alpha\chi^2\right),
\end{equation}
whose solution can not be associated with a first-order
differential equation. However, the solution of the equation of
motion connecting the vacua $M_3$ and $M_4$ is given by

\begin{equation}
\label{SNBPS} \chi(z)= \frac{m}{\sqrt{\lambda\alpha}}\tanh(\tilde
m z), \quad \tilde m = \sqrt{\frac{\alpha}{\lambda}}m,
\end{equation}
so, in this case, the fluctuation Hessian potential on the
$M_{34}$ wall becomes

\begin{equation}
\label{PNBPS} V_{NBPS}(z)=2m^2\left(
\begin{array}{cc}
\frac{2\alpha}{\lambda}-(1+\frac{2\alpha}{\lambda})sech^2(\tilde m z)&0 \\
0&\frac{\alpha}{2\lambda}-\frac{3\alpha}{\lambda}sech^2(\tilde m
z)
\end{array}\right).
\end{equation}
Note that as $\alpha\neq\lambda$ the tension of $M_{34}$ is
different from the $M_{12}$ wall tension. But, if
$\alpha=\lambda=m,$ we obtain

\begin{equation}
T_{34}=\int_{-\infty}^{\infty}\left(\frac 12(\chi^{\prime})^2+
V(0,\chi)\right)dz=m^2\int_{-\infty}^{\infty}sech^4(mz)dz= \frac
43 m,
\end{equation}
which is equivalent to the tension $T_{12}.$ In the limit
$\alpha<< \lambda,$ we have $\tilde m<<m,$ for very large range of
$z, V_{NBPS}\sim \hbox{diag} (-2m^2,0),$ which is constant and
non-positive. Therefore, the corresponding Hessian in Eq.
(\ref{PNBPS}) must have negative energy states, so that in this
range for the parameters $\alpha$ and $\lambda,$ the non-BPS
domain wall is not stable.

\subsection{BPS Solutions}

If we choose $\gamma=\frac{m^2}{2\lambda}$ in the trial orbit
above, we obtain the elliptical orbit

\begin{equation}
\frac{\lambda^2}{m^2}\phi^2+\frac{\lambda^2}{2m^2}\chi^2=1
\end{equation}
providing the following pair of BPS solutions

\begin{eqnarray}
\label{bps}
\phi(z)&&=\frac{m}{\lambda}
\tanh(\frac m2 z)\nonumber\\
\chi(z)&&=\pm\sqrt{2}\frac{m}{\lambda}sech(\frac m2 z),
\end{eqnarray}
for $\beta=\frac{\lambda}{2}$ and $\alpha=\frac{\lambda}{4}.$

 In
this case the four vacua become: $(\pm\frac{m}{\lambda}, 0)$ and
$(0,\pm\frac{2m}{\lambda}).$ It is easy to show that these static
two-field solutions of the BPS conditions are satisfied by the
equations of motion given by Eq. (\ref{E20}).

This pair of BPS solutions represents a straight line segment
between the vacuum states $M_1$ and $M_2$. The superpotential, in
terms of the bosonic fields, leads to the correct value for a
Bogomol'nyi minimum energy, corresponding to the BPS-saturated
state. Then, we see that, at classical level, according to Eq.
(\ref{Ebogo}), one may substitute

\begin{equation}
E_B^{min}=\mid W [\phi(z), \chi(z)]_{z=+\infty}- W [\phi(z),
\chi(z)]_{z=-\infty} \mid =\frac{4m^3}{3\lambda^2},
\end{equation}
which corresponds to the tension $T_{12}$ of the $M_{12}$ domain
walls. Indeed, some tensions of the $M_{ij}$ domain walls are
identical. Thus, the non-zero tensions of the domain walls are
$T_{13}=T_{14}=T_{23}=T_{24}=\frac 12
T_{12}=\frac{2m^3}{3\lambda^2}$. Note that these three walls
satisfy  a Ritz-like combination rule

\begin{equation}
T_{12}=T_{23}+T_{31},
\end{equation}
which is a stability relation generally seen in N=2 SUSY. One has
used the tension given by $T_{ij}=\mid W_{ij}\mid$ which is called
the Bogomol'nyi mass bound.

The projection of the potential over the $\phi$ axis provides the
double-well potential $V(\phi,
0)=\frac{\lambda^2}{2}(\phi^2-\frac{m^2}{\lambda^2})$ for the
kink-like domain walls given by Eq. (\ref{bps}).

In what follows we will discuss the stability of the two-field BPS
solutions and how the Lie graded algebra in SUSY QM is readily
realized via the factorization method. Note that, for the
potential model considered in this work, according to Eq.
(\ref{EV}), we can readily arrive at the following elements of
fluctuation Hessian matrix, $V_F(z):$

\begin{eqnarray}
\label{PE} V_{F12}(z)=V_{F21}(z)\equiv\frac{\partial^2}{\partial
\chi\partial\phi}V&&=
\frac{\partial^2}{\partial\phi\partial\chi}V=4
\alpha(2\alpha+\lambda)\phi\chi,\nonumber\\
V_{F11}(z)\equiv\frac{\partial^2}{\partial\phi^2}V&&=
6\lambda^2\phi^2-2m^2+2\alpha(2\alpha+
\lambda)\chi^2,\nonumber\\
V_{F22}(z)\equiv\frac{\partial^2}{\partial\chi^2}V&&=
6\alpha^2\chi^2+2\alpha(2\alpha+ \lambda)\phi^2- \frac{2\alpha
m^2}{\lambda}.
\end{eqnarray}

Thus, from the Ricatti equation (\ref{ER}) associated to the
non-diagonal flutuaction Hessian $V_{F}(z)$ given by Eq.
(\ref{PE}), we obtain

\begin{equation}
\label{E34a} \hbox{{\bf W}}(z)=-2\lambda\left(
\begin{array}{cc}
 \phi & \frac 14\chi \\
 \frac 14\chi & \frac 14\phi
\end{array}\right)_{\vert\phi=\phi(z),\chi=\chi(z)},
\end{equation}
where we have used the BPS state conditions (\ref{cbps}).
Therefore the bosonic zero-mode becomes

\begin{equation}
\label{mz} \Psi_-^{(0)}(z)=\Psi_{0}(z)=N\left(
\begin{array}{cc}
sech^2(mz)\\
sech^{\frac{2\alpha}{\lambda}}(mz)
\end{array}\right),
\end{equation}
which satisfies the annihilation condition $A^-\Psi_-^{(0)}(z)=0.$
The Eqs. (\ref{mz}) and (\ref{E34a}) are correct only for BPS
states, with $\alpha=\frac{\lambda}{4}.$

For the sector $\chi=0$, and $z\rightarrow\pm\infty$,
$m^2_\phi=4m^2$. Indeed, a possible soliton solution occurs even
when we choose $\beta=\lambda$ and $\gamma=\frac{m^2}{\lambda}$,
so that it implies  $\chi=0$ with a BPS domain wall, $M_{12}$,
which belongs to the topological soliton of the $\phi^4$ model,
whose solution is

\begin{equation}
\phi(z)=\frac{m}{\lambda}\tanh(mz).
\end{equation}

In this sector of the BPS states, the fluctuation potential
Hessian becomes

\begin{equation}
\label{POF} V_{BPS}(z) =m^2\left(
\begin{array}{cc}
6\tanh^2(mz)-2&0 \\
0&\frac{2\alpha}{\lambda^2}(2\alpha+\lambda)\tanh^2(mz)-
\frac{2\alpha}{\lambda}
\end{array}\right).
\end{equation}
In this particular case, the $\chi=0$ and $\phi $ = kink (or
anti-kink) configurations are the only BPS domain wall, so that
$W(z)$ becomes diagonal:

\begin{equation}
\label{EPBPS} \hbox{{\bf W}}(z)=-2\left(
\begin{array}{cc}
m\tanh(mz)& 0 \\ 0 & m\frac{\alpha}{\lambda}\tanh(mz)
\end{array}\right)_{\vert\phi =\phi(z), \chi =0}.
\end{equation}

Indeed, the bosonic sector of $H_{SUSY}$ is exactly given by
$O_F={\cal A}^+{\cal A}^-$, which, as obtained from the stability
Eq. (\ref{EE}), has the following zero-eigenmode:

\begin{equation}
\label{AF}
  \Psi_{-1}^{(0)}=\left(
\begin{array}{cc}
c_{(1)}sech^2(mz) \\
 0
\end{array}\right), \quad \Psi_{-2}^{(0)}=\left(
\begin{array}{cc}
0 \\
  c_{(2)}sech^{\frac{2\alpha}{\lambda}}(mz)
\end{array}\right),
\end{equation}
 $c_{(i)} being, i=1, 2$ the normalization constants of the
corresponding ground state. Note that, if $\alpha=\lambda$, we see
that the two SUSY representations of the superpotential diagonal
component become equivalent. One should also remark that $\alpha$
and $\lambda$ have been chosen both positive, thus
$\frac{2\alpha}{\lambda}>0$, so that as $\xi^{(0)}(z)$ is a
normalizable configuration.

We can see that $V_{BPS}$  is a diagonal Hermitian matrix, then
$O_F$ is also Hermitian. Hence, the eigenvalues $\omega^2_n$ of
$O_{F11}$ and $\tilde\omega^2_n$ of $O_{F22}$ are all real. In
this case, the fluctuation operator is diagonal, so we have two
representations of SUSY in quantum mechanics.  We shall explicitly
show that $\omega_n{^2}$ are non-negative, the proof of which
takes us to a solution of the P\"oschl-Teller potential
\cite{Morse}. Indeed, the mode equations are decoupled and may be
of two kinds, given by

\begin{equation}
\label{E38a}
 O_{F11}\eta_n\equiv
-\frac{d^2}{dz^2}\eta_n-m^2(6sech^2(mz)-4)\eta_n=\omega^2_n \eta_n
\end{equation}
and

\begin{equation}
\label{E39a} O_{F22}\xi_n\equiv
 -\frac{d^2}{dz^2}\xi_n-\frac{2m^2\alpha}{\lambda^2}(2\alpha+\lambda)
 sech^2(mz)\xi_n+m^2(
 \frac{2\alpha}{\lambda^2}(2\alpha+\lambda)-\frac{2\alpha}{\lambda})\xi_n=
 \tilde{\omega}^2_n\xi_n.
\end{equation}
Note that, according to Eqs. (\ref{EV}) and (\ref{DP}), if
$\alpha=0$, the potential becomes $V(\phi)=
\frac{\lambda^2}{2}(\phi^2-\frac{m^2}{\lambda^2})^2$, so the
stability equation is given by Eq. (\ref{E38a}) and, therefore,
there exists only the wall $M_{12}$.

Now, we can see that both types of solutions exist only for
certain discrete eigenvalues. Let us perform the transformation,
$mz=y,$ so that, upon comparison with the equation (12.3.22) in
\cite{Morse}, we obtain the following eigenvalues:

\begin{equation}
\label{E40a} \omega^2_n=m^2\left\{4-\left[\frac 52- (n+\frac
12)\right]^2\right\}.
\end{equation}
In this case,  we find only two bound states associated with the
eigenvalues $\omega^2_0=0$ and  $\omega^2_1=3m^2$ and, therefore,
the BPS states are stable.

Similarly, for the second type of solutions, we obtain, from Eqs.
(\ref{E39a}) and (12.3.22) of Ref. \cite{Morse}, the following
eigenvalues:

\begin{equation}
\label{E40b}
\tilde\omega^2_n=m^2\left\{\frac{2\alpha}{\lambda^2}(2\alpha+\lambda)
-\frac{2\alpha}{\lambda}-\left[\sqrt{\frac{2\alpha}{\lambda^2}
(2\alpha+\lambda)+\frac 14}-(n+\frac 12)\right]^2\right\}.
\end{equation}
In this case, we find the number of bound states given as below:

\begin{equation}
n=0,1, \cdots < \sqrt{\frac{2\alpha}{\lambda^2}
(2\alpha+\lambda)+\frac 14}-\frac 12,
\end{equation}
from which, we get

\begin{equation}
\label{E40c}
\tilde\omega^2_n=m^2\left\{16-(4-n)^2\right\},n=0,1,2,3,
\end{equation}
if we take $\alpha=2\lambda.$

We can see that, in this particular case, we have four bound
states associated with the eigenvalues $\tilde\omega^2_0=0$,
$\tilde\omega^2_1=7m^2$, $\tilde\omega^2_2=12m^2$ and
$\tilde\omega^2_3=15m^2$. Also, the eigenvalue for the zero mode
associated with the second type of solutions is given by

\begin{equation}
\label{E2mz}
\tilde\omega^2_0=m^2\left\{\frac{2\alpha}{\lambda^2}(2\alpha+\lambda)
-\frac{2\alpha}{\lambda}-\left[\sqrt{\frac{2\alpha}{\lambda^2}
(2\alpha+\lambda)+\frac 14}-\frac 12\right]^2\right\},
\end{equation}
which is non-negative if we assume

$$
\frac{\alpha}{\lambda} >
\frac{1}{\sqrt{2}}\left(\sqrt{\frac{2\alpha}{\lambda^2}
(2\alpha+\lambda)+\frac 14}-\frac 12\right).
$$
 Therefore we can conclude that the BPS state is stable.

\section{Conclusions}

\paragraph*{}

In the present work, we investigate, in terms of fluctuation
operators, the BPS domain wall states from a 2-dimensional model.
The corresponding stability equations have been analyzed for both
cases with and without supersymmetry. We have seen that domain
walls associated with the two-field potentials have features that
are not in the one-field models. However, if $\phi$ is a kink or
anti-kink configurations and $\chi=0,$ the second equation in
(\ref{ps}) implies $\beta=\lambda=\gamma=m,$ so only this
topological sector provides the interpolation between the (1,0)
and (-1,0) vacua, which corresponds $M_{12}$ BPS domain walls.

Also, we see that the elliptic orbit providing domain walls have
internal structure, so that the $\chi$ mesons are outside the host
defect and the $\phi$ mesons are inside. We present a particular
pair of BPS defect that is mapped into an elliptic orbit, whose
solutions map the Ising and Bloch walls in magnetic systems
\cite{IM}.

The connection between the (0,1) and (0,-1) vacua is realized the
via non-BPS solutions, for $\alpha=\lambda=m$. In this case, both
tensions $T_{34}$ and $T_{12}$ are equivalent. Our analysis of the
Hessian (\ref{NBPS}) shows us that the non-BPS domain wall is not
stable.

 The stability
equations associated with the soliton solutions of a simple model
of two coupled real scalar fields  have been investigated, by
calculating the tensions of the domain walls. In all cases, the
domain walls  belonging to the BPS states, the zero-mode ground
states become two-component eigenfunctions, for the particular
case in which we have found the explicit form for the situations
with different eigenvalues.

The result derived in this work for the matrix superpotential
given by Eq. (\ref{E34a}) in supersymmetry in Quantum Mechanics
(SUSY QM) is valid only for BPS solutions with $\alpha=\lambda/4.$

Also, results on sn-type elliptic functions given in Ref.
\cite{monsta}, for which boundary conditions on bound energy
levels of a classical system defined by one single scalar field,
and an extension to a relativistic system of two coupled real
scalar fields in a finite domain in (1+1) dimensions have been
investigated \cite{jorge99,jorge2}. According to our development,
we can readily realize the SUSY QM algebra, in coordinate
representation, for a 2-dimensional potential model in a finite
domain.

\vspace{1cm}

\centerline{\bf Acknowledgments}

We wish to thank CBPF-MCT and CFP-UFCG of Cajazeiras-PB, Brazil.
We would like to  acknowledge S. Alves for hospitality at CBPF of
Rio de Janeiro-RJ, Brazil, where the part of this work was carried
out and to J. A. Hela\"yel-Neto for many stimulating discussions
and criticisms on an earlier manuscript. This work was partially
supported by Conselho Nacional de Desenvolvimento Cient\'\i fico e
Tecnol\'ogico, CNPq.


\newpage


\unitlength=1cm
\begin{figure}[tbp]
\centering
\begin{picture}(10,1)
\epsfig{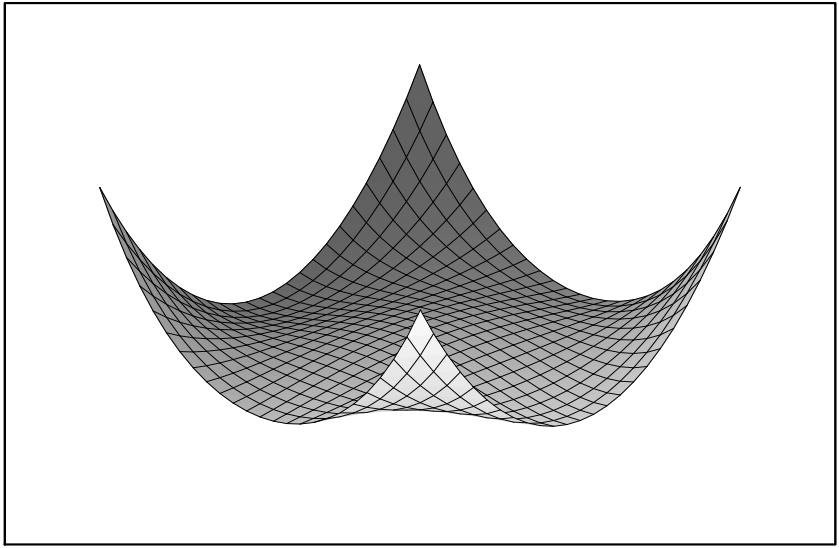}
\end{picture}
\vspace{7.5cm} \caption{ The two coupled field potential model
considered in this work.}
\end{figure}


\unitlength=1cm
\begin{figure}[tbp]
\centering
\begin{picture}(10,1)
\epsfig{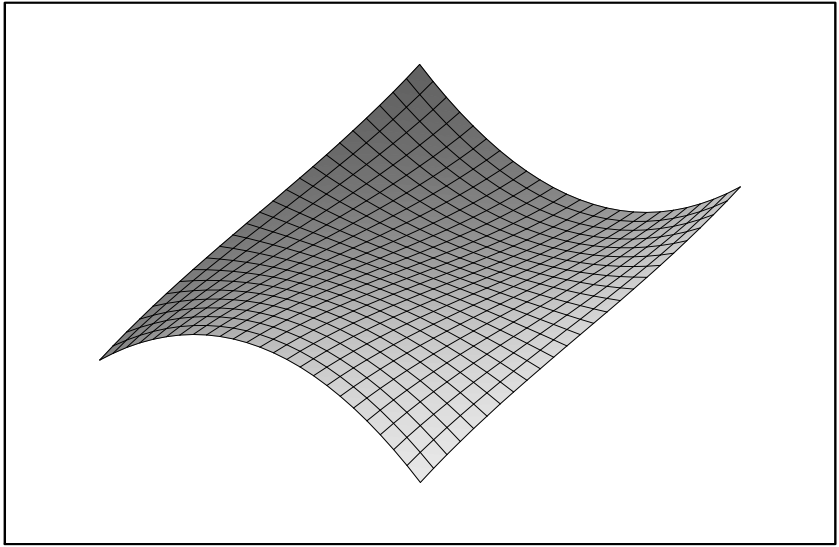}
\end{picture}
\vspace{7.5cm} \caption{The polynomial superpotential that
generates the BPS domain walls.}
\end{figure}

\newpage


\unitlength=1cm
\begin{figure}[tbp]
\centering
\begin{picture}(10,1)
\epsfig{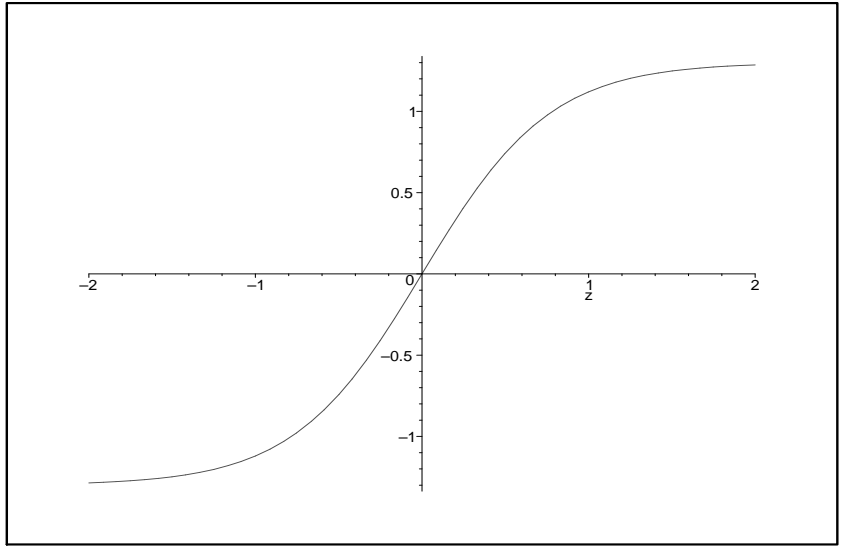}
\end{picture}
\vspace{7.5cm} \caption{ Static classical configuration which
represents the domain wall.}
\end{figure}


\unitlength=1cm
\begin{figure}[tbp]
\centering
\begin{picture}(10,1)
\epsfig{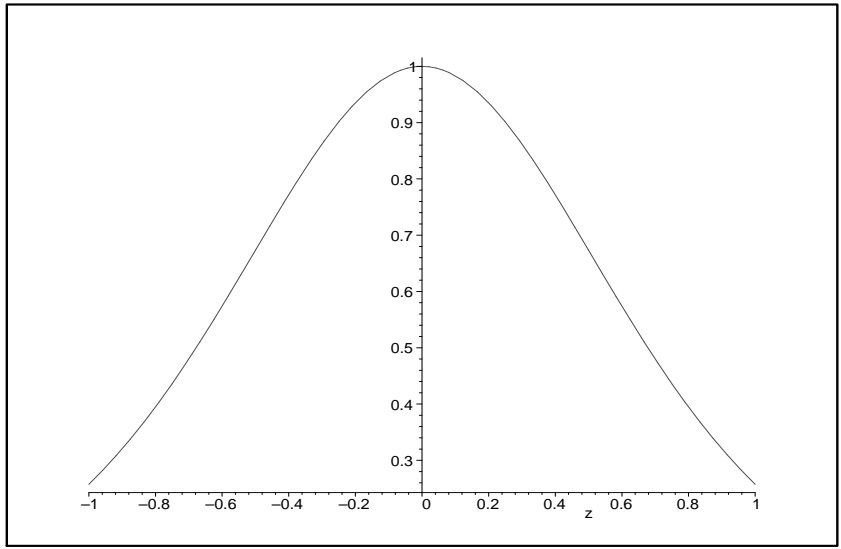}
\end{picture}
\vspace{7.5cm} \caption{ The eigenfunction of the zero mode, for
the topological sector $\chi=0$ and $\phi$=kink.}
\end{figure}

\end{document}